\documentclass[conference]{IEEEtran}
\IEEEoverridecommandlockouts
\usepackage{cite}
\usepackage{amsmath,amssymb,amsfonts}
\usepackage{graphicx}
\usepackage{textcomp}
\usepackage{xcolor}

\usepackage{algorithm}
\usepackage{algpseudocode}%
\usepackage{tabularx}
\usepackage{float}
\usepackage{tikz}
\usepackage{amsmath}
\usepackage{cuted, nccmath}
\usepackage{lipsum}
\usepackage{cite}
\usepackage{filecontents}
\usepackage{multirow}
\usepackage{float}
\usepackage{graphicx,subcaption}
\usepackage{enumitem}
\setitemize{noitemsep,topsep=0pt,parsep=0pt,partopsep=0pt}
\usepackage{lipsum}
\usepackage{titlesec}
\usepackage[width=\columnwidth]{caption}
\usepackage{tabularx}
\usepackage{algorithm}
\usepackage{amsmath}

\titlespacing*{\section}
{0pt}{1ex plus 0.2ex minus .2ex}{0ex plus 0.2ex minus .2ex}
\titlespacing*{\subsection}
{0pt}{1ex plus 0.2ex minus .2ex}{0ex plus 0.2ex minus .2ex}
\titlespacing*{\subsubsection}{0pt}{0ex plus 0.2ex minus .2ex}{0ex plus 0.2ex minus .2ex}
\titlespacing{\section}{0pt}{0pt}{0pt} 

\usepackage{authblk}
\begin{document}

\title{Interference and Need Aware Workload Colocation in Hyperscale Datacenters
}

\author[1,2]{Sayak Chakraborti}
\author[2]{Brian Coutinho}
\author[1]{Sandhya Dwarkadas}
\author[2]{Parth Malani}
\author[2]{Bikash Sharma}

\affil[1]{University of Rochester, Department of Computer Science, Rochester, NY, USA}
\affil[2]{Meta Platforms Inc., Menlo Park, CA, USA}

\affil[ ]{Emails:\textit{ \{schakr11,sandhya\}@cs.rochester.edu, \{bcoutinho,pmalani,bsharma\}@fb.com}}

\maketitle

\begin{abstract}
\textcolor{black}{
Datacenters suffer from resource utilization inefficiencies due to the conflicting goals of service owners and platform providers.  Service owners intending to maintain Service Level Objectives (SLO) for themselves typically request a conservative amount of resources. Platform providers want to increase operational efficiency to reduce capital and operating costs. 
Achieving both operational efficiency and SLO for individual services at the same time is challenging 
due to the diversity in service workload characteristics, resource usage patterns that are dependent on input load, heterogeneity in platform compute, memory, I/O, and network architecture, and resource bundling (e.g., compute and memory on a single node).
}

\textcolor{black}{
This paper presents a tunable approach to resource allocation that accounts for both dynamic service resource needs and platform heterogeneity. \textcolor{black}{In addition, an online K-Means-based service classification method is used in conjunction with an offline sensitivity component.}
Our tunable approach allows trading resource utilization efficiency for absolute SLO guarantees based on the service owners' sensitivity to its SLO. 
We evaluate our tunable resource allocator at scale in a private cloud environment with mostly latency-critical workloads.  When tuning for operational efficiency, we demonstrate up to $\sim$50\% reduction in required machines; $\sim$40\% reduction in Total-Cost-of-Ownership (TCO); and $\sim$60\% reduction in CPU and memory fragmentation, but at the cost of increasing the number of tasks experiencing degradation of SLO
by up to $\sim$25\% 
compared to the baseline.  When tuning for SLO, by introducing interference-aware colocation, we can tune the solver to reduce tasks experiencing degradation of SLO  by up to $\sim$22\% 
compared to the baseline, but at an additional cost of $\sim$30\% in terms of the number of hosts. We highlight this trade-off between TCO and SLO violations, and offer tuning based on the requirements of the platform owners.
}
\end{abstract}


\section{Introduction}

\label{introduction}

\begin{figure}[h!] 
\includegraphics[scale=0.7]{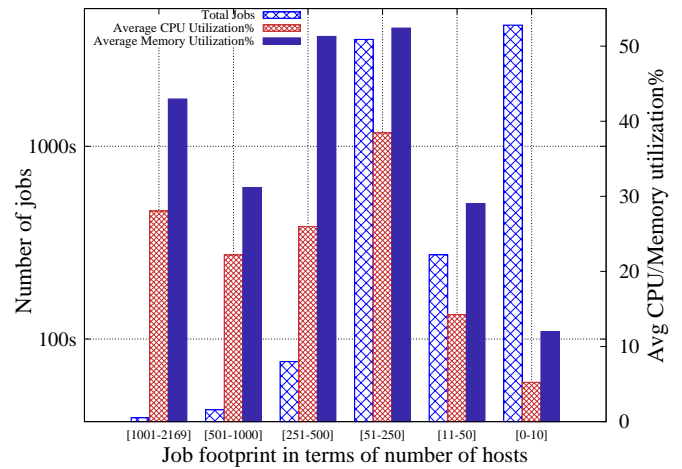} 
\caption{Average CPU/memory utilization of a \textcolor{black}{production} task pool based on hosts allotted (tasks are grouped/binned based on number of hosts utilized)}
\label{AvgAds}
\end{figure}

There are many challenges in managing a hyperscale datacenter that runs thousands of jobs and machines, and incurs large capital and operating expenditures (CapEx).
Among them, improving resource efficiency, which can reduce CapEx, while maintaining the service level objective (SLO\footnote{\textcolor{black}{Service Level Indicators (SLIs) are service-level metrics, for example, p99 latency, while the Service Level Objective (SLO) is the limit on the SLI, for example, the p99 latency being less than 1 ms. 
}}), is a non-trivial problem to solve. 
In cloud platforms such as Google, Alibaba and AWS~\cite{Amazon,Alibaba,Google},  compute and memory resources are severely underutilized (the average CPU utilization being $\sim$20\%-40\% and memory being $\sim$40\%). 
Figure~\ref{AvgAds} shows the current CPU/memory utilization for one of the job pools in our organization. 
During normal operation, the average CPU utilization across our fleet is $\sim$35-40\%. Around 93\% of the jobs (each utilizing less than 250 hosts) have on average 20\% CPU utilization and 31\% memory utilization. Thus, there exists a long tail of jobs with low resource utilization: a perfect opportunity for stacking.
Ensuring the highest amount of resource utilization is also paramount for overall energy efficiency of a hyperscale datacenter~\cite{Energy-Proportional} with measurable impact on the environment. 

In order to meet the SLO of a service\footnote{We use the term service/microservice/workload interchangeably.} owner, resources are often over-provisioned \textcolor{black}{(pessimistic allocation)}. Without careful management, over-provisioning can result in varying degrees of under-utilization of the allocated hardware resources (compute, memory, network, flash, etc.) over time.
Additionally, contingencies in planning for right-sized data centers such as disaster recovery, host failures, and service growth, also contribute to over-provisioning of resources. Multi-tenancy or colocation (running two or more jobs on the same server or host) offers opportunities to reduce the under-utilization resulting from over-provisioning, but relies on dynamic resource managers and load balancers to handle any abnormal spikes in resource usage.

Colocation requires consideration of both specified resource requirements as well as actual usage patterns and hardware diversity to avoid SLO violations. Workload characterization helps understand actual usage patterns and their resource demands and bottlenecks. Workload characteristics of different services tend to be diverse, with differing resource bottlenecks. Colocation policies can benefit from this information by colocating tasks that are not bottlenecked on similar resources. 

However, despite finding candidates with complementary resource requirements, colocated tasks can interfere with each other, thereby hampering each other's performance (resulting in an SLO violation) due to contention for shared resources that are not partitioned or isolated, such as last level cache (LLC), memory bandwidth, or network bandwidth. While techniques for partitioning shared resources like the  LLC~\cite{CAT} and memory bandwidth~\cite{RDT} are available, they incur additional overhead, need reconfiguration on every host reassignment, and are difficult to configure at scale. 

Static policies using a one-time classification of applications based on resource utilization or service category (latency-critical vs. batch) have been proposed~\cite{Bubble-Up, PerfIso, Stretch}. In contrast to prior assumptions of typical server workloads, our fleet has a 9:1 ratio for latency-critical vs. batch workloads; thus, colocation based on complementary workload category results in under-utilization due to the inability to find colocatable workloads. 
Additionally, resource utilization characteristics are diverse and vary dynamically based on periodic traffic patterns, datacenter type (public vs. private), kind of deployment (bare-metal vs. VMs\cite{ResourceCentral}), and over time with application changes.

Hyperscale datacenters generally tend to have diverse hardware fleets caused by the limited multi-year life cycle of racks and a preference for purchase of the latest hardware to reduce Total Cost of Ownership (TCO), resulting in architectural heterogeneity. Heterogeneity in machine architectures in terms of their capacity, specifications, and cost add to the challenges of efficient workload colocation. For example, if a compute heavy job specifies the need for 5 CPU cores to execute, choosing one over another machine may result in differing actual utilization due to differences in their CPU architecture. 

To address these challenges, we propose a tunable approach to resource allocation based on low-overhead online workload resource usage characterization and offline sensitivity analysis. 
The online workload characterization technique dynamically learns workload resource usage patterns by periodically measuring actual resource usage. In order to increase operational efficiency, we select candidates with orthogonal resource requirements for colocation so that they avoid interfering with each other.
The offline sensitivity analysis technique uses candidate services to quantify sensitivity to interference on shared resource dimensions. 
 Integration of shared resource sensitivity in colocation policies offers an advantage of not having hard partition limits on shared resources. The characterization and sensitivity analysis are combined in a platform-aware manner to account for heterogeneity in server architectures. The user/service owner can remain platform-agnostic  since the runtime takes care of adapting to heterogeneity in the server architecture.

\textcolor{black}{Our experimental setup uses $\sim$37,000 tasks and $\sim$18,000 hosts, showing the scalability of our approach. Our methodology does not rely on workload classification into latency-critical and batch, or on their specific SLO metrics, and can be applied to colocate two or more jobs of any type without violating their individual service-level objectives (SLOs). Our objective function can be tuned to achieve up to $\sim$50\% reduction in hosts required, reduce TCO by $\sim$40\%, and reduce CPU and memory fragmentation by $\sim$60\% when using percentile 99 resource utilization based limits, but increases SLO violations by $\sim$25\% relative to the baseline. By incorporating sensitivity scores, the SLO violations are reduced by $\sim$22\% relative to the baseline, but at the cost of an increase of $\sim$30\% in host count. Our techniques offer platform owners the flexibility to tune this cost vs SLO trade-off depending on their objectives.}
 
The key takeaways of our work are summarized below:
\begin{itemize}
      \item Resource requirement specification by service owners is very conservative, experimentally observed using isolated experiments. 
      \item An online multi-resource workload characterization technique that uses the $99^{th}$ percentile resource utilization is able to minimize loss of SLO. 
      \item  An offline method to derive sensitivity scores for a given (job, resource) pair, normalized to account for server heterogeneity, is able to avoid over-provisioning and improve workload colocation efficiency.
      \item  Combining the offline sensitivity scores with the online workload characterization into an objective function for resource allocation and mapping allows tuning the system to trade SLO against resource efficiency.
\end{itemize}

\section{Related work }
\label{related work}



\textbf{\textcolor{black}{Workload classification:}} 
Many works~\cite{Heracles, PARTIES,CLITE, Stretch,Quasar, PerfIso} categorize workloads into latency-critical and batch, and develop policies that prioritize jobs based on such classification.  
Heracles\cite{Heracles} proposes the usage of resource slack available at low-load levels for latency-critical applications to run batch workloads. They use a feedback-based controller to colocate batch tasks with latency-critical services by dynamically adjusting multiple software and hardware isolation limits.
PARTIES~\cite{PARTIES}  and CLITE~\cite{CLITE} expand to allowing multiple latency-critical workloads on the same host. 
PARTIES uses a state machine to explore a fixed subset of configurations/resource allocations with hardcoded priorities, resulting in potentially suboptimal decisions for some applications. CLITE learns the appropriate resource allocation and colocation across workloads by using a small subset of resource configurations to train a bayesian optimization model. 
PerfIso\cite{PerfIso} reserves a small percentage of cores within a host that can be used by latency-critical workloads as and when required for load fluctuation. Batch workloads run opportunistically when resources are available and are not provided any progress guarantees. The scalability of these online mechanisms to 1000s of jobs is yet to be explored. 

\textcolor{black}{Autopilot\cite{Autopilot} and Resource Central\cite{ResourceCentral} are resource management methods at large-scale datacenters (at Google Cloud and Azure, respectively) that use dynamic feedback-based machine learning techniques to determine appropriate resource allocation in order to meet SLOs while improving resource utilization.
}




Most of these approaches dynamically learn the behavior of workloads and adjust resource allocation to meet SLOs. However, the limitations include: 1) The focus is to meet the SLOs of latency-critical workload(s) at the expense of a best-effort/batch workload. \textcolor{black}{Based on our observations, batch workloads constitute less than 10\% of our fleet's total number of tasks, thus reducing the opportunity for colocation of one or more batch workloads with a single latency-critical workload. Also, as microservice-based application architectures gain popularity, it has become challenging to distinguish jobs on a single dimension of latency-critical versus batch.} Hence, an ideal classification and mix of latency and batch workloads is non-trivial. Batch workloads also have throughput-oriented metrics that need to be maintained.  2) The iterative search to find optimal resource configurations proposed in \cite{PARTIES,CLITE} doesn't hold good in a dynamic environment faced with constant fluctuations in load and underlying infrastructure. \textcolor{black}{ 3) Machine learning-based approaches such as the ones in \cite{ResourceCentral, Autopilot} often require a huge training set of resource configurations that is not readily available and have limited exploration points. Hence there arises a need to generate synthetic training sets.} 4) Lastly, these approaches optimize for resource slack locally but do not address how to pick the right mix of workloads for colocation to improve overall fleet utilization.

\textbf{Sensitivity analysis:}  
Quasar\cite{Quasar} uses a combination of offline workload characterization
and profiling along heterogeneity, interference, and scale up/out dimensions. 
Information on sensitivity (interference) to colocation is maintained in a 2-dimensional form  containing pairwise sensitivity information. In contrast, we maintain a per-workload sensitivity score that is independent of the colocated workload (but captures the impact of stress on investigated resource dimensions). 

Bubble-Up\cite{Bubble-Up} uses a tunable microbenchmark  to characterize performance degradation/sensitivity to contention for a resource by adjusting the pressure on the resource (bubble). 
\textcolor{black}{Bubble-Flux\cite{Bubble-Flux} addresses the shortfalls of the static profiling techniques used by Bubble-Up, particularly, the inability to adapt to input and load changes, 
by introducing a Dynamic Bubble mechanism (real-time and instantaneous compared to a controlled environment) and an Online Flux Engine. The Online Flux Engine dynamically tracks SLI for the latency-sensitive application and accordingly throttles the execution of the colocated batch workload.}

\textcolor{black}{PYTHIA~\cite{pythia} discusses the issues of colocating multiple batch workloads with a single latency-critical workload on the same host. Previous approaches\cite{Bubble-Up, Bubble-Flux, Quasar} use an additive sum of the sensitivity metric of batch workloads to make decisions on colocation. The authors show that a trained linear regression using a small subset of configurations can predict contention more precisely than a simple additive model. 
} 

\textcolor{black}{Our methodology for sensitivity analysis resembles that of Bubble-Up\cite{Bubble-Up, Bubble-Flux} but avoids the learning cost of determining per application pair sensitivity curves.
Instead, we perform an offline sensitivity analysis on candidate services and store a sensitivity score that is a threshold based on the service's SLO. Since the threshold is based on the service's SLO, it incorporates the service's sensitivity to interference across all resources in the system: CPU, network, and all levels of the memory hierarchy.   
} 



\textbf{Server heterogeneity aware mechanisms:}
Server heterogeneity is an important aspect of deciding efficient resource utilization. 
\textcolor{black}{\cite{Google,Quasar} address server heterogeneity in datacenters by either offline resource characterization of server architectures using benchmarks or quantifying application performance based on exploration (collaborative filtering based).}
We use an offline microbenchmark based relative/normalized resource characterization along all resource dimensions for heterogeneous server architectures.


\textbf{Microarchitecture-based mechanisms:}
Some works (\cite{DataSharing}, \cite{CoherenceStalls}) use hardware performance counters to detect and alleviate performance degradation caused by data sharing and memory bandwidth saturation in NUMA architectures. \textcolor{black}{Other works~\cite{Dirigent, CacheSlicershahrad2021provisioning} leverage existing microarchitectural mechanisms like DVFS and Cache partitioning to manage QoS at fine scales based on workload classification and competition for resources. CacheSlicer~\cite{CacheSlicershahrad2021provisioning} allocates cache partitions to VMs in a public cloud platform using a classification of VMs into 3 categories based on their sensitivity to competition in the LLC. Dirigent~\cite{Dirigent} uses cache partitioning and frequency control to improve the performance of foreground latency-critical jobs while also minimizing variance in performance.} Others (e.g., Stretch\cite{Stretch}) propose changes to the microarchitecture, such as partitioning the ROB (Reorder Buffer) entries asymmetrically while colocating latency-sensitive and batch workloads together on SMT (Simultaneous Multi-Threading) cores. 
\textcolor{black}{
Other approaches include boot-time or dynamically configurable knobs (like core/uncore frequency, LLC ways partitioning, types of prefetchers, etc.) \cite{SoftSKUs} and usage of accelerators \cite{Accelerometer} based on workload characterization to improve application performance. 
These approaches can be used in a complementary fashion with our policy for potentially improved isolation.
}

\section{Interference and Need-Aware Colocation}
\label{methodology}
The main components of our proposed technique are characterization and clustering based on resource utilization (Section~\ref{Characterization}), server heterogeneity (Section~\ref{diversity of server hosts}), sensitivity analysis of candidate services from the clusters (Section~\ref{sensitivity}), and interference-aware colocation using an in-house integer-linear-programming packer-solver (Section~\ref{rebalancer}). 

\textcolor{black}{A workload/service can comprise one or more jobs (logical units). A job is a collection of tasks running on possibly multiple hosts. A task is a single running process on a single host.} 
\textcolor{black}{Platform-wide profiling is available using performance counters\cite{intel_developer_manual}. Performance metrics are attributed to running tasks/jobs using the perf\cite{perf_event_open} subsystem and are stored in a database that can be queried based on jobs (aggregated) or tasks as required.}

Workload re-characterization based on resource utilization, as well as colocation/relocation using the packer-solver, is done online periodically (every 1 hour in our experiments). This allows workloads with changes in resource usage patterns to migrate between clusters and thus enables the packer-solver to use current resource usage limits and sensitivity scores for packing. 

Sensitivity analysis is performed offline on individual tasks belonging to a job by determining the impact on performance under different resource pressures (see Section~\ref{sensitivity}). 
Candidate jobs
mostly belong to mature services that rarely migrate to other clusters. The resulting sensitivity scores are assigned to individual tasks and are used during the packing process. 
\subsection{Resource utilization based workload characterization}
\label{Characterization}
We use the percentile resource usage of dimensions like CPU utilization, memory footprint utilization, and network bandwidth utilization\footnote{Metrics are collected by the kernel per container} collected per job/task averaged over 60 seconds. The percentile function sorts the resource usage in ascending order and checks for the $n^{th}$ percent sample, and returns the associated value. In a nutshell, the percentile gives us a limit beneath which $n$ percent of the observations were made. For example, percentile 99 of a particular resource dimension means the resource usage was below this limit for 99 percent of the observations made.


We use a conservative approach of using percentile 99 (discussed in Section~\ref{setting limits}) resource utilization limits and use K-means clustering technique to group tasks into clusters based on their percentile resource usage. Clustering gives us important insights into resource utilization and trends of tasks in clusters. Clustering information can also be used for setting resource limits when onboarding new services.

\subsection{Server Heterogeneity}
\label{diversity of server hosts}
The server fleet consists of a diverse range of server architectures across both generations and product lines (Table ~\ref{Architectures}). These architectures differ in microarchitecture, core count, topology, memory hierarchy and capacity, and memory and network bandwidth. This diversity adds to the complexity of job placement. 
To quantify the throughput/performance each server architecture
is capable of, we use in-house throughput scores measured using representative benchmarks. The scores capture the microarchitectural differences across architecture types, as well as the impact of core counts, memory hierarchy and capacity, and cpu topology. These scores allow us to translate resource need and sensitivity (see Section~\ref{sensitivity}, where we also discuss sensitivity to network and memory bandwidth) measured on one architecture to another.  \textcolor{black}{The representative benchmarks are derived to mimic the characteristics of industrial workloads deployed in our datacenters and are engineered to stress multiple resource dimensions.} 
\vspace{0.2cm}
\begin{center}
\begin{table}[tbh!]
\resizebox{1\linewidth}{!}{ \begin{tabular}{||c | c | c | c | c|c ||} 
 \hline
 Architectures (server type) & Cores & Memory (in GB) & MemBw (in GBps) & NetBw (in Gbps)& Score \\ [0.5ex] 
 \hline\hline
Haswell10 (Type I) & 48 & 32 & 75.6 & 5&1.63x  \\ 
 \hline
Haswell12 (Type II)& 48 & 256 & 120 & 5&1.79x  \\
 \hline
Skylake14 (Type I)& 36 & 64 & 76 & 25&1.29x\\
 \hline
Skylake16 (Type II)& 80 & 256 & 170 & 12.5&2.52x\\
 \hline
Broadwell18 (Type I)& 32 & 32 & 42 & 25&1x\\ 
\hline
Broadwell20 (Type II)& 56 & 256 & 153 & 12.5&1.98x\\ [1ex] 
\hline
\end{tabular}}
\caption{Comparison between different architectures 
}
\label{Architectures}
\label{Costs}
\end{table}
\end{center}
\vspace{-0.65cm}

To limit the complexity, we categorize the server architecture into two umbrella types labeled here as \textbf{Type I} and \textbf{Type II}. We perform workload clustering in a type-specific manner and use normalization when moving tasks from one type to another. \textbf{Type I} hosts have smaller compute and memory capacity and are less costly when compared to \textbf{Type II} hosts, which have more compute and memory resources. \textcolor{black}{Services with higher resource demands are generally allocated \textbf{Type II} hosts.} Table~\ref{Costs} shows those differences and also compares architectures within each umbrella type. The scores have been normalized to Broadwell18, which has the fewest cores and memory. In order to minimize the TCO, we want to minimize resource usage not just based on the number of hosts being utilized but also based on the type of hosts. 

\subsection{Sensitivity Analysis}
\label{sensitivity}
Sensitivity analysis is performed for modeling and quantifying the performance impact (using the key metric(s) that comprise the SLI)
of the injected interference or restricted usage of particular resource dimensions. We assign numeric values, which we term as $sensitivity \hspace{0.1cm} scores$ that are architecture and resource dimension specific. \textcolor{black}{We acquire a tier of homogeneous hosts (10 or more) to run the target job for these experiments.}
\label{sensitivity experiments}
\textcolor{black}{} 
We use A/B tests and in-house load-testers to perform these experiments. A/B tests check two variants of the same service environment but with a different experimental setup. For example, the A test could be running without any interference/restriction while the B test would run with added interference/restriction. \textcolor{black}{Load-testers have the capability to tune traffic load to desirable limits. All experiments are performed at maximum load limits. When load-testers are not available, we use shadow traffic, which replays production traffic as input load}. 
We primarily target three resource dimensions for our study: compute, memory bandwidth, and network bandwidth.  
\begin{itemize}
\item \texttt{Compute:} We restrict the number of cores utilized by the service (using taskset) to measure the impact of restricted cores. 
\item \texttt{Memory Bandwidth:} We use Intel Memory Latency Checker (MLC)\cite{MLC} to inject excess memory bandwidth and memory usage.
\item \texttt{Network Bandwidth:} We use iperf\cite{iperf} to add extra network bandwidth on hosts and see the impact it has on services running on it.
\end{itemize}
\textcolor{black}{We use  CPU task affinity to limit the compute resources utilized by MLC\cite{MLC} and iperf\cite{iperf} for our experiments and keep an eye on memory usage. While these resources can't be decoupled entirely, we make sure that the workload in question gets adequate resources as requested. Taking memory capacity and bandwidth into account accommodates the impact on other memory hierarchy components like shared caches.}

The \textbf{sensitivity score} is a real number/fraction assigned for the particular host's architecture and a specific resource dimension. The sensitivity score is a threshold of resource availability below which the application can experience SLO degradation. `One' represents 100\% of the particular resource dimension needs. 
In other words, the application should be guaranteed this fraction of the shared resource for normal operation. A lower score indicates that the workload can tolerate usage/contention of that particular resource for colocation, and this usage is indifferent to the key metrics of the service. A higher score suggests that the application's key metric is sensitive to this particular resource dimension and can't be compromised. We change the  fraction of contention for a resource dimension and check its impact on the key metric(s). 

As suggested earlier, we perform these experiments using A/B tests and load-testers running on specific hosts with a specific architecture. The $sensitivity \hspace{0.1cm} scores$ collected is specific to the architecture it is collected on. To use these scores for other architectures, we normalize the scores based on the target architecture's resource capacity. We use equation~\ref{CPU sensitivity normalization} for sensitivity normalization across architectures. The intuition behind the equations is the resource disparity among architectures. Primarily, differences in micro-architectures and memory/network bandwidth per core are taken into account for normalization. \textcolor{black}{ We use per-core bandwidth capacity for memory and network bandwidth capacity normalization. The cores are normalized based on relative performance using synthetic benchmarks which stress the core performance (refer Section~\ref{diversity of server hosts}). Upon normalization, the sensitivity score may exceed the value `one' for hosts with lower resource capacity. Thus, these smaller hosts won't be able to accommodate such services.}

\begin{equation}\label{CPU sensitivity normalization}
\begin{matrix}
CPU Sens_{Arch1} = \frac{CPU Sens_{Arch2} \times Score_{Arch2}}{Score_{Arch1}}\\
MEMBW Sens_{Arch1} = \frac{MEMBW Sens_{Arch2} \times \frac{MEMBW_{Arch2}}{Cores_{Arch2}}}
{\frac{MEMBW_{Arch1}}{Cores_{Arch1}}}\\
NETBW Sens_{Arch1} = \frac{NETBW Sens_{Arch2} \times \frac{NETBW_{Arch2}}{Cores_{Arch2}}}
{\frac{NETBW_{Arch1}}{Cores_{Arch1}}}
\end{matrix}
\end{equation}

As it is infeasible to perform sensitivity analysis on all tasks, we make a simple assumption. \textcolor{black}{} We assume that \textit{tasks in the same cluster behave similarly}. We assign the sensitivity scores collected from candidate services for each cluster to all tasks in the same cluster. These architecture and task-specific sensitivity scores help the packer-solver colocate tasks while ensuring minimum SLI degradation.
\subsection{Packer-solver: Bringing it all together}
\label{rebalancer}
We use an in-house packer-solver for efficient colocation in our task pool.
The packer-solver runs periodically, taking task pool and host pool as inputs. We inject percentile 99 resource usage limits and sensitivity scores, which are task, architecture, and resource dimension specific, into the packer-solver. 
The packer-solver (constrained optimizer with objective function) is an assignment solver that can use a variety of bin-packing algorithms. It assigns objects (tasks) to containers (hosts). Each task is assigned to precisely one host while satisfying some goals and constraints. The packer-solver starts with an arbitrary assignment and performs simple valid moves to improve objectives defined by goals and constraints. Goals are defined metrics that are converged toward, and multiple goals can be combined linearly using weights. Constraints, on the other hand, are independent properties/limits that need to be held. 

\begin{equation}
\resizebox{1.00\hsize}{!}{
$
\begin{split}
\label{Goals2}
h \in (Hosts Occupied \cup Hosts Free), k \in [CPU, MEMORY]\\
h_{\textbf{limit}}(h,k) = host\hspace{0.1cm}resource \hspace{0.1cm}capacity\hspace{0.1cm} for\hspace{0.1cm}  resource\hspace{0.1cm}k\\
t_{\textbf{limit}}(t,k) = task\hspace{0.1cm}resource \hspace{0.1cm}requirement\hspace{0.1cm} for\hspace{0.1cm} resource\hspace{0.1cm}k\\
tasks(\textbf{h})= tasks\hspace{0.1cm} assigned\hspace{0.1cm} to\hspace{0.1cm} host\hspace{0.1cm} h\\
Fragmentation(k) = \sum_{\substack{\tiny \forall h\in Hosts Occupied}}(h_{\textbf{limit}}(h,k) - (\sum_{\substack{\forall t \in tasks(\textbf{h})}}(t_{\textbf{limit}}(t,k)))  )\\
\textbf{Constraint:} \forall h \in Hosts Occupied, \forall k, h_{\textbf{limit}}(h,k) >= (\sum_{\substack{\forall t \in tasks(\textbf{h})}}(t_{\textbf{limit}}(t,k)))\\
\textbf{Goal:} Minimize(Hosts Occupied)\\
\textbf{Secondary Goal:}
Minimize(\sum_{\substack{\tiny \forall k}}Fragmentation(k))\\
\forall \textit{h} \in \textit{Hosts Occupied}, \hspace{0.3cm}\exists \textit{t} \in \textit{TASKS},\\
\textbf{Soft constraint:}\sum_{\substack{t}}Sensitivity(t , h) <= 1\\
\end{split}
$
}
\end{equation}

\textcolor{black}{One such goal of the solver is to minimize the number of hosts utilized (and thereby TCO) and the resource fragmentation, such that each task is mapped to exactly one host and task resource requirements are met (constraints). Heterogeneity in architecture during placement is taken care of by normalizing sensitivity scores to the  underlying architecture. Equation~\ref{Goals2} shows the goals for convergence. As the number of hosts, tasks, and resource counts (like the number of cores, memory capacity) cannot be fractions, the constrained optimization problem is solved using Integer Linear Programming using a greedy approach to converge to a result as quickly as possible.   
}

Additional goals include reducing SLO violations, achieved by colocating tasks based on their $sensitivity\hspace{0.1cm} scores$. The linear sum of the sensitivity score for any particular resource dimension on the target host should not exceed 100\% in order to prevent violation of SLO of the scheduled tasks. 
Equation~\ref{Goals2} depicts our goal formally. Here, \textit{Sensitivity(t, h)} represents the sensitivity of  task \textit{t} on that particular host \textit{h}.

\section{Evaluation Methodology}
\label{sec:methodology}

Our problem scope consists of more than 1000 production jobs with $\sim$37,000 tasks and $\sim$18,000 hosts.
\textcolor{black}{
Table~\ref{Service Characterization} depicts some of the production workloads used for sensitivity analysis. Log1 is a real-time joiner to prepare training data for ranking models. Also, it keeps an in-memory storage structure to maintain some maps. KeyVal1 is a heavily optimized in-memory key-value store for storing dense features for ranking and is memory-bound. NN1 performs Nearest-Neighbour search in vector space with millions of vectors. KeyVal2 and KeyVal3 are key-value stores where KeyVal2 is customized to store user data, and KeyVal3 is heavily optimized for in-memory storage of dense features for ranking and is compute-bound. Rank2 is a ranking model to find and deliver the best-suited ads as output. These workloads differ from SPEC2006\cite{SPEC2006} and Google traces\cite{ProfilingGoogleTraces, MemoryHierarchyOfWebSearch} as described in \cite{SoftSKUs, Accelerometer}. Most of these production workloads are well-tuned for performance, where multiple performance metrics are collected at regular intervals.
}

\begin{center}
\begin{table}[h!]
\resizebox{\columnwidth}{!}{ \begin{tabular}{||c ||c ||c ||}  \hline
 Service & Category & Cluster \\ [0.5ex] 
 \hline\hline
Log1 & Logging & low  \\ 
 \hline
KeyVal1  & In-Memory Key-Value Store (memory bound) & low\\
 \hline
NN1  & Nearest-Neighbor(NN) search & medium \\
 \hline
KeyVal2 & Customized Key-Value store & medium\\
 \hline
KeyVal3 &In-Memory Key-Value Store (compute bound) & high\\
\hline
Rank2  & Ranking & high\\ [1ex] 
\hline
\end{tabular}}
\caption{Service characterization}
\label{Service Characterization}
\end{table}
\end{center}

\vspace{-1cm}
\section{Results}
\label{results}
\subsection{Workload characterization parameters}
\label{workload characterization results}

\subsubsection{\textcolor{black}{Number of workload clusters for K-Means clustering}}
\begin{figure}[h!] 
 \centering
\includegraphics[width=\columnwidth,height=7cm]{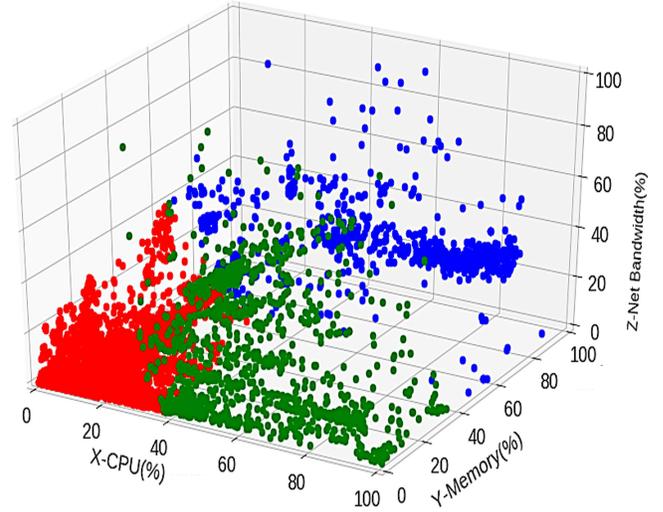}
\caption{Resource utilization clustering (for Type II hosts) based on measured p99 resource utilization along 3 dimensions (i) CPU, (ii) memory usage, and (iii) network bandwidth 
}
\label{clusters}
\end{figure}

\textcolor{black}{
We used an off-the-shelf statistical method to determine the appropriate number of workload clusters. 
Within Cluster Sum of Square (WCSS) shows the compactness of the points within a cluster. The lower the sum, the better the representation. 
}
From our observations, the elbow of the curve is around the value "three," so we use three as the number of clusters. Figure~\ref{clusters} depicts the percentile 99 resource utilization-based clusters. Tasks from the low resource utilization cluster (ones in red) are good candidates for colocation with tasks in other clusters or among themselves. 
Candidate services can then be chosen from each cluster to perform sensitivity analysis. Table~\ref{Service Characterization} shows different candidate services from our clusters.

\subsubsection{Setting percentile limits}
\label{setting limits}
We evaluate the impact on the SLI metric(s) and SLO violations of workloads based on the percentile resource usage limits guaranteed. 
\textcolor{black}{ We vary both CPU and memory limits, but illustrate the CPU percentile limits in the figures as memory limits remain close to original requested allocation.}

\begin{figure}[h!] 
\includegraphics[scale=0.7]{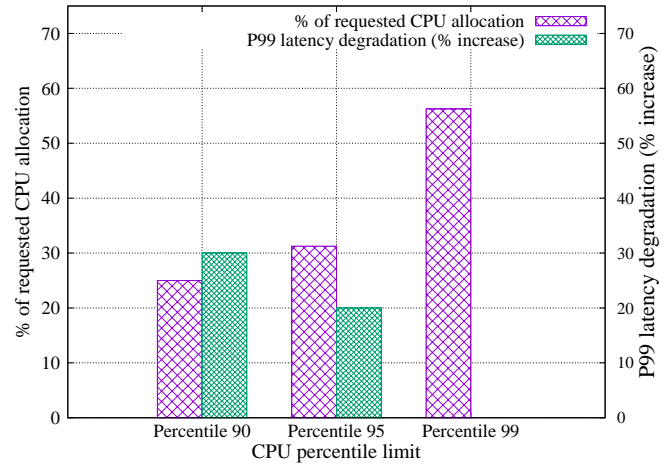}
\caption{KeyVal2 performance with changing resource 
requirement limits compared to requested CPU allocation (on Skylake16) 
}
\label{CPU percentile for service KeyVal2}
\end{figure} 
\begin{figure}[h!]
\includegraphics[scale=0.7]{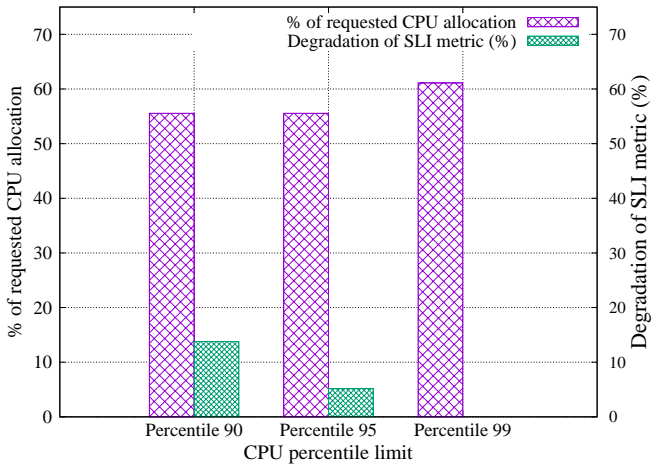}
\caption{Log1 performance with changing resource requirement limits compared to requested CPU allocation (on Skylake16) 
}
\label{CPU percentile for service Log1}
\end{figure} 
Figures~\ref{CPU percentile for service KeyVal2} and~\ref{CPU percentile for service Log1} show the impact of setting different CPU percentile limits for the services KeyVal2 and Log1 (Table ~\ref{Service Characterization}). 
The CPU percentile limit is used to control the actual CPU allocation to a workload to be a fraction of the requested original CPU allocation (represented on the left-hand Y axis). CPU percentile utilization is determined by taking CPU utilization measurements every 1 second, averaging over a minute, and taking the percentile utilization over 7 days. We observe that for KeyVal2, reducing the CPU resource allocation impacts the P99 latency (percentile 99 server latency measured from end to end, a key metric for this service). \textcolor{black}{We make similar observations for Log1 and other workloads. Note that the percentile 99 CPU allocation for KeyVal2 is $\sim$55\% of the originally requested CPU allocation. We use percentile 99 limits to determine resource allocation when tuning for operational efficiency.}
Thus, we can recover unused resources for better resource utilization and colocation. 
\subsection{Incorporation of Sensitivity}
\label{sensitivity analysis results}
\begin{figure}[h!] 
\includegraphics[scale=0.7]{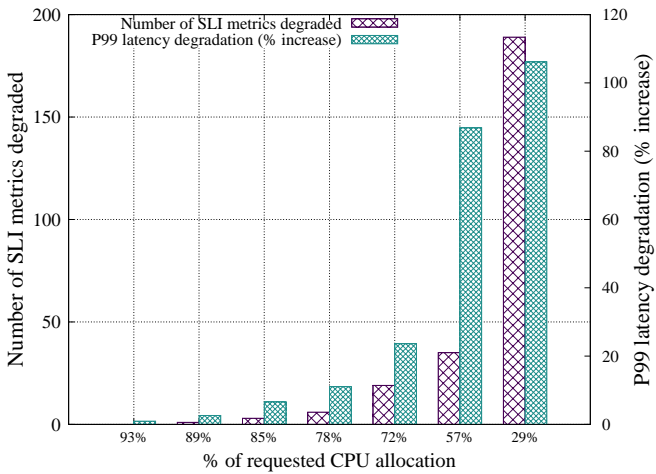}

\caption{CPU sensitivity for Rank2 (on Broadwell20)}
\label{Adfinder CPU}
\end{figure} 

\textcolor{black}{ 
We evaluate the impact of changing the percentage of CPU resources allocated relative to the requested allocation at a finer granularity. 
We measured the sensitivity for six candidate services, two from each cluster (Table~\ref{Service Characterization}). 
}

\textcolor{black}{ Degradation in SLIs can cause SLO violations. Services often use multiple SLI metrics such as goodness of ranking, throughput, etc., with varying priorities. 
Figure~\ref{Adfinder CPU} shows the impact of reducing compute resource allocation (as a fraction of the requested allocation) on service Rank2, plotting P99 latency as well as the total number of SLIs experiencing degradation (an SLI is noted as experiencing degradation when it sees $>$ 5\% change). 
} \textcolor{black}{We observe that Rank2's SLI metrics degrade in a nonlinear fashion as the percentage of requested compute cores actually allocated is decreased. Based on our observation, we set the CPU sensitivity value to 0.93 as at 93\% of requested CPU allocation, we see less than a 5\% impact on its SLIs.} 

For memory bandwidth sensitivity experiments, we use two variants, i) we restricted the number of cores used for bandwidth injection, and, ii) no restriction on cores. We also correlated the shared LLC misses (L3 for most architectures) with the effects on the key metric(s) to confirm the interference caused due to contention on shared resources.
\textcolor{black}{We observe that when we restrict the interference causing application to a fraction of the total cores, the interference experienced by NN1 is much less and has a lower impact on its key metric. Thus, we conclude that
containerization using partitionable resources helps reduce interference by controlling use of the unpartitionable resources.}

\begin{figure}[h!] 
\includegraphics[scale=0.7]{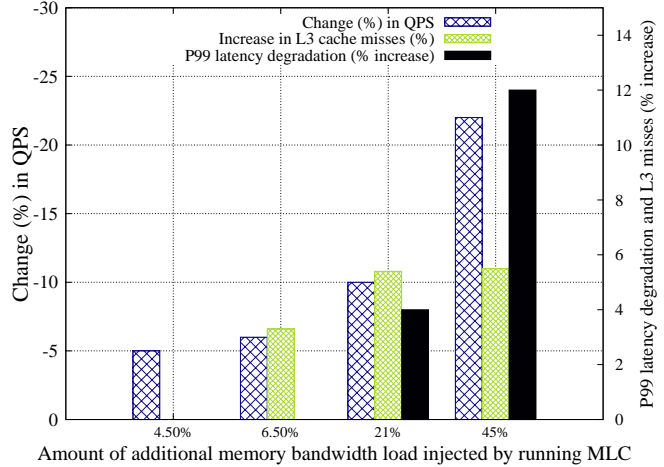} \caption{Impact of injected Memory bandwidth on service NN1 (10 percent cores utilized for MLC)(on Haswell10)}
\label{Mem bandwidth for service NN1 (10 percent)}
\end{figure}
\begin{center}
\begin{table}[htp!]
\resizebox{\columnwidth}{!}{ \begin{tabular}{||c | c| c | c | c | c |c ||} 
 \hline
 \multirow{2}{*}{\hspace{-0.1cm}Service }&\multicolumn{3}{c|}{Haswell10}&\multicolumn{3}{c||}{Broadwell20}\\
    &  CS& MBS& NBS& CS& MBS& NBS\\
 \hline\hline
{\scriptsize Log1} &0.83 &0.92 &1.32 &0.68 &0.6 &0.6 \\
\hline
{\scriptsize KeyVal1} &0.77 &0.63 &1.32  &0.63 &0.41 &0.61 \\
\hline
{\scriptsize NN1} &0.77 &\textbf{0.79} &0.8 &0.63 &0.53 &0.36\\
\hline
{\scriptsize KeyVal2} &0.75 &0.92 &1.12 &0.64 &0.45 &0.51 \\
\hline
{\scriptsize KeyVal3} &0.99 &1.15 &1.98 &0.81 &0.76 &0.9\\
 \hline
{\scriptsize Rank2} &1.13 &1.31 &1.98 &\textbf{0.93} &0.86 &0.96\\
 \hline
\end{tabular}}
\caption{\textcolor{black}{CPU(CS), Memory Bandwidth(MBS) and Network Bandwidth(NBS) Sensitivity Scores for services listed in Table~\ref{Service Characterization}. Sensitivity scores are determined experimentally on one architecture and derived using Equation~\ref{CPU sensitivity normalization} on other architectures.}}
\label{SensitivityTables}
\end{table}
\end{center}
\vspace{-1cm}
\textcolor{black}{Using p99 latency as the SLI, based on Figure~\ref{Mem bandwidth for service NN1 (10 percent)}, we assign NN1 a memory bandwidth sensitivity score of 0.79 as at 21\% of interfering bandwidth, the impact on p99 latency was around 7\%. Reducing the impact to 5\% would require a sensitivity score between 0.79 and 0.935, a data point not represented in the Figure~\ref{Mem bandwidth for service NN1 (10 percent)}.} 

\subsection{Packer-solver results}
\label{packer solver results}
We ran experiments using a packer-solver on a representative workload consisting of more than 1000 production jobs with $\sim$37,000 tasks and $\sim$18,000 hosts. Since we are unable to directly influence production workloads, the packer-solver was executed independently/offline using a trace from a live production pool as input and producing the changes in allocation and mapping as output.

Hosts in the system are occupied or freed due to task movement by the packer-solver. 
\textcolor{black}{In order to limit the time spent in the packer-solver, we limited the maximum number of moves the packer-solver was allowed to make to 400. Setting a limit on the total number of task moves also control over the adverse effects of the cost of task movement.}



\begin{center}
\begin{table}[h!]
\resizebox{\columnwidth}{!}{ \begin{tabular}{||c | c | c | c | c |c ||} 
 \hline
 Config & CPU limit & Mem limit & CPU sen & MemBw Sen & NetBwSen\\ [0.5ex] 
 \hline\hline
Baseline & original & original & no & no & no\\  
\hline
P99 CPU & P99  & original & no & no & no\\  
\hline
P99 Mem & original & P99 & no & no & no\\  
\hline
P99 & P99 & P99 & no & no & no\\  
\hline
P99 Sens& P99 & P99 & yes & yes & yes\\[1ex] \hline
\end{tabular}}
\caption{Configurations used for Packer-solver experiments}
\label{configuration}
\end{table}
\end{center}
\textcolor{black}{In order to tune for operational efficiency, we allow colocation based on the percentile 99 (\textbf{P99}) resource utilization of workloads so as to make use of unutilized resources within an allocation. In order to tune for a workload's SLO, we add sensitivity scores to effect interference-aware colocation (refer Equation~\ref{Goals2}) (\textbf{P99 Sens}).
The tunable component of our approach is the weight attached to the sensitivity goals. Assigning smaller weights is more beneficial in terms of reducing host counts but at the cost of probable SLO violations.} 
Table~\ref{configuration} lists the configurations used in the runs for our in-house packer-solver.
\begin{figure}[!h] 
  \centering
\includegraphics[scale=0.7]{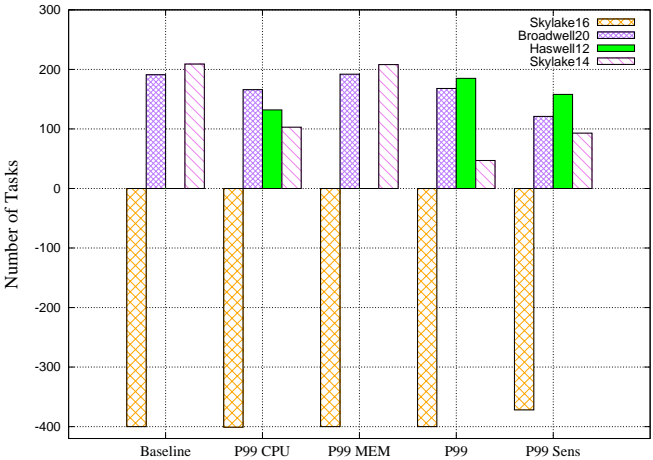} 
\caption{Tasks Moved for a total moves of 400 (configuration from Table~\ref{configuration})} 
\label{fig:400 jobs}
\end{figure}
\begin{figure}[!h]
  \centering
\includegraphics[scale=0.7]{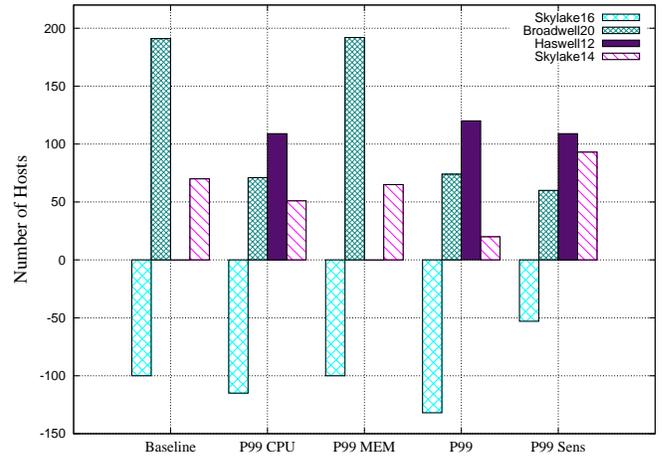} 
\caption{Hosts Moved for a total moves of 400 (configuration from Table\ref{configuration}) } 
\label{fig:400 hosts}
\end{figure}

Figures~\ref{fig:400 jobs} and~\ref{fig:400 hosts} show the tasks and host changes for a maximum of 400 total task moves by the packer-solver for different configurations (Table~\ref{configuration}). We observe that using percentile 99 limits
helps free up more 
\textit{Skylake16} hosts (which have a higher score (refer Table~\ref{Architectures})) by moving tasks to other architectures that can accommodate the tasks. 

\begin{figure}[!h] 
  \centering
\includegraphics[scale=0.7]{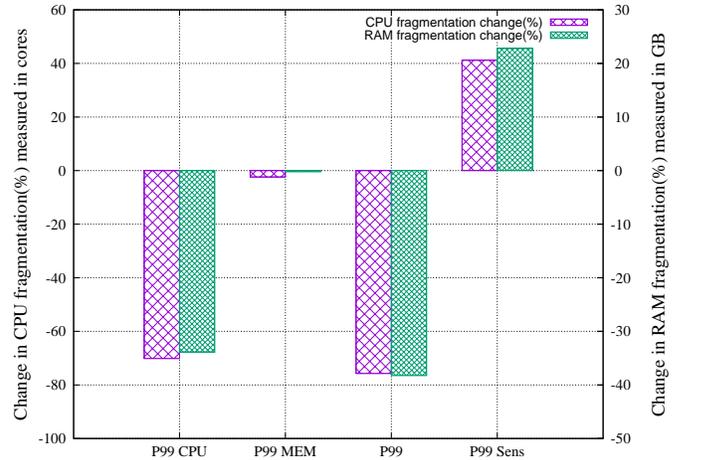} 
\caption{Fragmentation change compared to baseline (configuration from Table\ref{configuration}), lower is better} 
\label{fig:400 frags}
\end{figure}

Figure~\ref{fig:400 frags} shows the resultant fragmentation of CPU cores and memory (RAM) \textcolor{black}{(refer Equation~\ref{Goals2})} for each configuration, measured in terms of the percentage of unoccupied cores within occupied hosts. \textcolor{black}{Compared to the baseline, the \textbf{P99} configuration reduces hosts utilized by $\sim$50\% and \textbf{P99 Sens} increase hosts utilized by $\sim$30\%.} Resource fragmentation is defined as the remainder of resource capacity that is not utilized on an occupied host. We observe that using combined percentile 99 resource utilization for CPU and memory reduces fragmentation significantly compared to the  baseline. The fragmentation increases when we introduce sensitivity scores as expected since the baseline doesn't incorporate sensitivity information (for the baseline, the placements are based on limits set by service owners).

\textcolor{black}{Quantifying the reduction in TCO compared to baseline, \textbf{P99} configuration is able to reduce Type II server cost by 31.86\%, Type I cost by 71.42\%, and reduce overall TCO cost by 39.94\%.
} 
The cost is derived based on tasks and hosts movement for 400 total moves and is normalized against the baseline. Freeing up more \textbf{Type II} reduces the TCO significantly given the ratio of cost of a \textbf{Type I} to a \textbf{Type II} server. \textcolor{black}{Reducing the TCO helps in capacity planning as well, as requirement for costly hosts reduces. 
}

\begin{figure}[h!] 
  \centering
\includegraphics[scale=0.7]{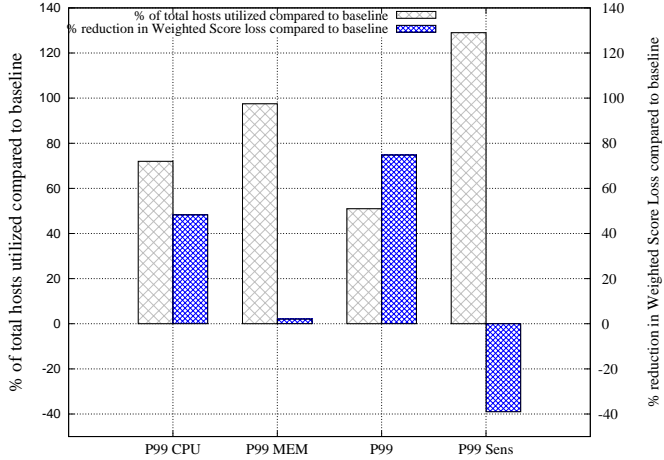} 
\caption{Weighted-Score-Loss (WSL) change compared to baseline (configuration from Table\ref{configuration}), positive reduction and lower total hosts used is better} 
\label{fig:400 RCU}
\end{figure}

Based on architecture-specific relative scores (mentioned in Table~\ref{Architectures}), we also depict the Weighted-Score-Loss (WSL) and the total number of hosts utilized in Figure~\ref{fig:400 RCU}. \textcolor{black}{TCO is calculated based on the ratio of cost of the umbrella server type (\textbf{Type I} vs. \textbf{Type II}), while WSL uses the scores assigned to specific CPU architecture (refer Table~\ref{Architectures}).}
The  \textbf{P99} configuration reduces TCO, WSL significantly and also reduces the total number of hosts used. WSL is defined in Equation~\ref{WCL}.
\begin{equation}\label{WCL}
\resizebox{0.85\hsize}{!}{%
$
\begin{split}
WSL=\sum_{\substack{j}}((Score[j].Occupied[j])-(Score[j].Freed[j])),\\
\forall \textit{j} \in {Architectures},\\
\textit{Occupied}:No.\hspace{0.1cm} of \hspace{.01cm} hosts\hspace{0.1cm} occupied/used,\\
\textit{Freed}:No.\hspace{0.1cm} of \hspace{.01cm} hosts\hspace{0.1cm} freed/released
\end{split}
$
}
\end{equation}
\begin{figure}[!h]
  \centering
\includegraphics[scale=0.7]{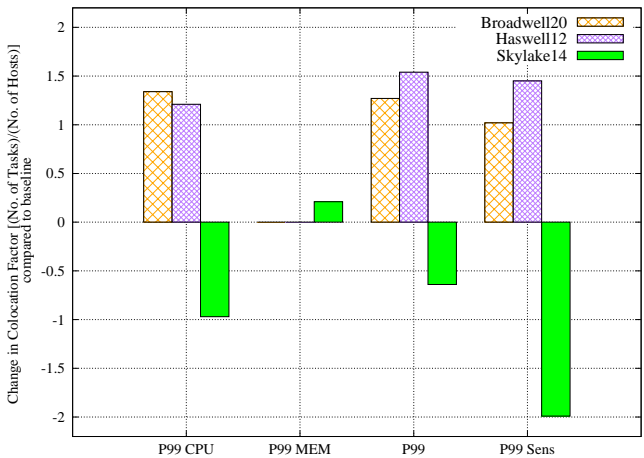}
\caption{Colocation factor compared to baseline (configuration from Table\ref{configuration}) } 
\label{fig:400 colocation efficiency}
\end{figure} 
Figure~\ref{fig:400 colocation efficiency} shows the colocation factor for each architecture. We observe better colocation factor for \textit{Broadwell20} in most configurations over the baseline, primarily due to the usage of percentile 99 limits. Also, as \textit{Broadwell20} has higher resource capacity, it can accommodate more workloads. The secondary cause for this observation is for hosts with higher resource capacity, the sensitivity scores are lower for most workloads, providing greater capacity for colocation. Similarly, for \textit{Skylake14}, which is a  \textbf{Type I} server, the colocation factor is lower due to limited room for colocation of sensitive tasks.

Comparing the \textbf{P99} configuration with the baseline shows up to 50\% reduction in hosts required, reduction in CPU and memory fragmentation by $\sim$60\%, and a reduction of  $\sim$75\% of WSL for this pool. We also reduce $\sim$40\% of the TCO for the same configuration. Freeing up costly hosts helps in the reduction of TCO and WSL. Configuration (\textbf{P99 Sens}) that involves sensitivity is unable to colocate tasks into \textbf{Type I} servers (as explained above) compared to the baseline, which causes degradation in both fragmentation and WSL. We quantify the gains of the sensitivity aware configuration in the following paragraphs.
\subsubsection*{\textbf{Quantifying interference}}
\label{QoS violation}

\textcolor{black}{
Finally, after bin-packing completes, we use sensitivity scores that are calculated based on changes in SLI metrics to check for SLO violations.
We speculate on the potential for SLO violations by using any infringement by the packer-solver's placement decisions on the fraction of resources required by the service (as implied by its sensitivity score). Any infringement is considered a violation. 
} 
\begin{figure}[H] 
\includegraphics[scale=0.7]{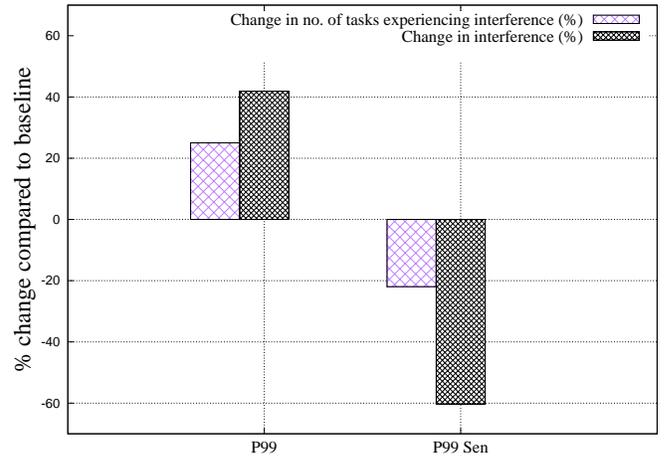} 
\caption{Quantifying likelihood of interference (-ve is better) }
\label{QoS}
\end{figure} 
We try to quantify the changes in interference that can degrade the SLI based on whether or not the packer-solver pays attention to sensitivity. We present two metrics: i) number of tasks that experience probable interference (tasks on hosts where the sum of sensitivity score is $>$1), and ii) interference quantified as the sum of the (portion of) sensitivity scores in excess of 1 (see Equation~\ref{interference}). We use our sensitivity lookup table (per task sensitivity scores for each dimension) to check the scores for each resource dimension and count violations on hosts based on the current placement scenario. Though this is a suggestive metric, in Figure~\ref{QoS}, we depict \textbf{P99} and \textbf{P99 Sens} configurations and show that using sensitivity scores reduces the chances of interference by a significant margin over the baseline. We show that with \textbf{P99 Sens} we are able to reduce up to $\sim$22\% of tasks experiencing SLI degradation compared to baseline. For \textbf{P99} configuration, there are $\sim$25\% more tasks that experience SLI degradation compared to the baseline. 

\begin{equation}
\begin{split}
interference=\sum_{\substack{h \in \textit{HOSTS}}}
\textit{if} \hspace{0.3cm}((\sum_{\substack{t \hspace{0.10cm} on\hspace{0.10cm} h}}\textit{Sensitivity(t , h)}) > 1)\\
((\sum_{\substack{t \hspace{0.10cm} on\hspace{0.10cm} h}}Sensitivity(t , h)) - 1)
\label{interference}
\end{split}
\end{equation}

\section{Discussion}
\label{Discussion}

\subsection{Granularity of metric collection}
\label{granularity}
The metrics collected for resource utilization averages over a one-second granularity samples for a minute. We evaluate if there are any gains in accuracy by using finer-grained sampling. 
There are overhead implications of storing metrics at a finer granularity. As we use percentile 99 resource usage information, we performed experiments to determine the impact of the metric collection granularity. The percentile 99 utilization computed at a finer granularity remains within 1\% of that at our default granularity for our workloads.

\subsection{Tackling spikes}
\label{tackling spikes}
Resource usage spikes for applications can arise due to change in input load or due to application phases. Since we calculate the percentile resource usage information based on 1 minute aggregated resource usage per task over seven days, we argue that we take care of resource fluctuations. We also use sensitivity scores measured across multiple dimensions for bin-packing. The sensitivity scores are assigned based on SLI metric(s) at near peak load. 

\textcolor{black}{In addition to this, there is an online mechanism in place to change resource allocation limits or preempt tasks if required. Techniques such as scale-up: provide higher resource limits/better resources; and scale-out: start multiple replicas of the same job to distribute work; can be used.}

\subsection{Dynamism of approach}
\label{Dynamism of approach}
As discussed earlier, our methodology has both offline and online components.  The sensitivity analysis occurs offline. Resource utilization-based characterization and bin-packing by the packer-solver take place at regular intervals (every 1 hour). This online approach allows the packer-solver to take updated resource usage into account, helping tasks with variable resource usage patterns get re-classified to a different cluster. The packer-solver is a low-overhead solver that has been deployed at scale to handle a vast number of tasks.

\subsection{Cost of task movement}
\label{job movement}
Moving tasks across hosts can be costly, but we can tune the number of tasks that should be moved and reduce the ill effects of task movement when not necessary. At scale, task managers have in-built constraints to refrain from making task movements that will severely impact network traffic. Also, there are redundancies based on replicas and restrictions based on the physical location of tasks as per datacenters.

\subsection{Overheads}
Most datacenters report utilization of various metrics for internal performance overview and assessment. Our methodology taps into these existing metrics.
The offline sensitivity analysis using load-testers/replicated production traffic is done periodically but at lower frequencies, usually once every few days. This is performed only on some selected candidate services. The overhead of sensitivity analysis is similar to other sensitivity based methods such as Bubble-Up\cite{Bubble-Up}. The packer-solver takes the percentile resource usage information and sensitivity scores as inputs and performs tasks to host placement periodically as is the norm. 

\section{Conclusion}
\label{conclusion}
\textcolor{black}{Given the heterogeneity of available resources and the varied characteristics and resource usage of workloads running in datacenters, efficient resource utilization is a challenging objective to achieve. In this work, we shed light on understanding and addressing this challenging objective at scale.}

We demonstrate that by characterizing and clustering workloads based on their percentile resource usage limits, we are able to combine this information with offline sensitivity and hardware heterogeneity analysis to effectively allocate and map resources to workloads.  Our technique eliminates the need for services to be classified into latency-critical or batch workloads.

Using our clustering technique, sensitivity analysis, and packer-solver, we can perform interference-aware colocation of workloads on hosts and thus,  i) Colocate tasks that won't hamper each others performance, and ii) reduce the need for additional provisioning of resources via more efficient utilization. We show that using our techniques, we can tune the packer-solver to achieve up to 50\% reduction in hosts required, reduce CPU and memory fragmentation by $\sim$60\%, reduce the TCO by $\sim$40\%, and WSL by $\sim$75\%. These savings come, however, with an increase in SLO violations by $\sim$25\% compared to the baseline.
Through the incorporation of sensitivity scores, we are able to reduce up to $\sim$22\% tasks experiencing interference relative to the baseline at an additional cost of $\sim$30\% in host count. This is the result of contradictory goals of resource efficiency vs maintaining service SLOs. Our methodology can be tuned to trade TCO for SLO violations based on the service owner's requirements.


\bibliographystyle{plain}
\bibliography{ref}
\end{document}